\documentclass[twocolumn,prb]{revtex4}
\usepackage{amsmath}
\usepackage{graphicx}
\usepackage{float}
\usepackage{amssymb}
\usepackage{times}
\DeclareGraphicsExtensions{.eps,.ps}

\newcommand{\Av}[1]     {\left\langle #1 \right\rangle}

\def\bi         {\begin{itemize}}
\def\ei         {\end{itemize}}
\def\benu	{\begin{enumerate}}
\def\eenu	{\end{enumerate}}
\def\bmat       {\left[ \begin{array}}
\def\emat       {\end{array} \right]}
\def\beq	{\begin{equation}}
\def\eeq	{\end{equation}}
\def\beqn       {\begin{eqnarray*}}
\def\eeqn       {\end{eqnarray*}}
\def\beqa       {\begin{eqnarray}}
\def\eeqa       {\end{eqnarray}}
\def\bquote	{\begin{quote}}
\def\equote	{\end{quote}}
\def\f          {\frac}
\def\bwide	{\begin{widetext}}
\def\ewide	{\end{widetext}}

\def\ve		{\varepsilon}

\def\w          {\omega}

\def\z		{\zeta}

\def\D          {\Delta}

\def\Si          {\Sigma}

\def\bk         {{\bf k}}

\def\bp         {{\bf p}}
\def\bq         {{\bf q}}

\def\dag	{\dagger}

\begin{document}

\title{Theory of non-Fermi liquid near a diagonal electronic nematic state
on a square lattice}

\author{Ying-Jer Kao}
\affiliation{Department of Physics and Center for Theoretical Sciences, National Taiwan University, Taipei, Taiwan 106}
\author{Hae-Young Kee}
\affiliation{Department of Physics, University of Toronto, Toronto,
Ontario, Canada M5S 1A7}

\date{\today}

\begin{abstract}
We study effects of Fermi surface fluctuations on a single-particle life time
near the diagonal electronic nematic phase on a two-dimensional square lattice.
It has been shown that there exists a quantum critical point (QCP) 
between the diagonal nematic and isotropic phases.\cite{Doh06prb}
We study the longitudinal fluctuations of the order parameter near the critical point, 
where the singular forward scattering leads to a non-Fermi liquid behavior over 
the whole Fermi surface except along the $k_x$- and $k_y$-directions. 
We will also discuss the temperature and chemical potential dependence of
the single-particle decay rate.
\end{abstract}
%74.20.-z Theories and models of superconducting state
%71.45.Gm Exchange, correlation, dielectric and magnetic response functions, plasmons
%71.10.Hf Non-Fermi-liquid ground states, electron phase diagrams and phase
%transitions in model condensed matter systems 
%71.27.+a Strongly correlated electron systems; heavy fermions

\maketitle

{\it Introduction} ---
In strongly correlated electronic materials, 
electrons can organize themselves into a pattern with periodicity
differing from that of the underlying lattice along a particular direction.
The phase with such a non-trivial charge ordering
breaks a discrete translational symmetry of underlying lattice along one direction,
and has been named as stripe.
The stripe physics has been of great interest to both theoretical and experimental condensed
matter physicists, especially because its quantum
fluctuation may play an important role in understanding
the mechanism of high temperature superconductivity.\cite{Tranquada95nature,Salkola96prl,Zaanen89prb} 
The clear evidence of such a novel paired stripe order was reported
by Mori et al in electron diffraction experiment on (La,Ca)MnO$_3$.\cite{Mori98nature}

It has been proposed that phases with fluctuating stripes can be generic ground states of 
a doped Mott insulator 
introduced by the hole doping.\cite{Kivelson98nature,Kivelson03rmp}
These intermediate forms of matter have been dubbed as electronic smectic and nematic phases.
While the smectic breaks a translational symmetry, the electronic nematic phase
breaks a discrete rotational symmetry of underlying lattice, which can be viewed
as fluctuating stripes. The electronic nematic phase has been
referred as Pomeranchuk instability of Fermi liquid,\cite{Pomeranchuk58jetp} and  has been 
studied in theoretical models such as t-J, (extended) Hubbard, and Fermi liquid models.\cite{Yamase00jpn,Halboth00prl,Hankevych02prb,Nilsson05prb}
The electronic nematic phase has been suggested as a phase responsible
for  novel phenomena observed in strongly correlated systems.
These phenomena include, for example,
the anisotropic resistivities in high Landau levels of quantum Hall systems,\cite{Lilly99prl} 
the anisotropic scattering patterns observed in one of high temperature superconductors,
\cite{Hinkov04nature,Kao05prb}
and the metamagnetic transition and anisotropic transports in Sr$_3$Ru$_2$O$_7$.\cite{Grigera04science,Grigera03prb,Kee05prb}

In addition to the anisotropic nature of electronic nematic phase, it was shown that
the gapless Goldstone mode of the nematic phase in continum model
leads to the non-Fermi liquid behavior in
single-particle scattering rate on Fermi surface except along the symmetric directions.\cite{Oganesyan01prb}
However, using the same model of quadrupole density interaction on the square lattice, 
it was shown that the putative continuous transition is 
preempted by a first order
transition due to the van Hove singularity. 
The regime of nematic phase diminishes exponentially as one approaches the continuum limit of small density with 
approximately circular Fermi surface.\cite{Kee03prb,Khavkine04prb}
Therefore, there is no critical Fermi surface fluctuations for the non-Fermi liquid behavior in 
realistic systems with underlying lattice; both transverse (phase)
and longitudinal (amplitude) modes are gapped.
On the other hand, it was later found that additional uniform 
interaction in the forward scattering channel suppresses the first order transition.\cite{Yamase06prb}
This restores the non-Fermi liquid behavior at the Fermi surface, except along 
the zone diagonal direction, 
via the coupling to the fluctuation of order parameter at the quantum critical point.\cite{DellAnna06prb}

In this paper, we present  theory of  non-Fermi liquid  near the diagonal nematic phase.  It is worthwhile to note that there are two distinct nematic 
states in the square lattice.\cite{Doh06prb}
Previous studies provide understanding of phenomena related to 
Fermi surface distortion along the parallel direction of $k_x$- or $k_y$-axis of the Brilluoin zone.
The diagonal nematic phase can be characterized by a Fermi surface distortion along the diagonal direction of the Brilluoin zone.  
It is shown that the transition between the diagonal nematic and isotropic phases
is second order, because it does not involve the splitting of the van Hove singularity in
the density of states.\cite{Doh06prb}
We find the non-Fermi liquid behavior in the single-particle life time at the Fermi surface
, through the coupling to the longitudinal mode of diagonal nematic order,
except along the zone parallel directions (x- and y-axis of Brillioun zone).
Our numerical results of self energy show the expected exponent of $2/3$ in frequency similar
to the parallel nematic phase\cite{DellAnna06prb} and the continuum model\cite{Oganesyan01prb}.
However, the physical origin of the non-Fermi liquid behavior is different for these two cases.

{\it Model} --- The model Hamiltonian for the parallel and the
diagonal nematic phases on a square lattice is written as follows,
\beqa \nonumber
\label{eq:hamiltonian}
H &=& \sum_{\bk\sigma} \ve_\bk c_{\bk\sigma}^\dag c_{\bk\sigma}^{} \\
\nonumber
& -& \sum_{\bk\bk'\bq\sigma \sigma^\prime} \left[
F_2(\bq) \z_1(\bk)\z_1(\bk')
+G_2(\bq) \z_2(\bk)\z_2(\bk')\right] \\
&&\times
c_{\bk+\f{\bq}{2}\sigma}^\dag c_{\bk'-\f{\bq}{2}\sigma^\prime}^\dag
c_{\bk'+\f{\bq}{2}\sigma^\prime}^{} c_{\bk-\f{\bq}{2}\sigma}^{},
\eeqa
where  $ \z_1(\bk)= \cos k_x-\cos k_y$, and 
$ \z_2(\bk)= 2 \sin k_x\sin k_y$ are the form factors originating from
the quadrupole density-density interaction\cite{Oganesyan01prb}, and $\ve_\bk$ is the tight binding dispersion 
$\ve_\bk=-2t(\cos k_x+\cos k_y)-\mu$.  
We assume short range quadrupolar interactions of
$F_2({\bf q})$ and $G_2({\bf q)}$ with the form of $F_2/(1+\kappa q^2)$ and $G_2/(1+\kappa q^2)$, 
respectively.

The order parameters associated with the parallel ($\Delta_P$) and diagonal ($\Delta_D$) 
nematic phases are given by
\beqa
\nonumber
\D_P &=& F_2 \sum_{\bk} (\cos k_x - \cos k_y)\Av{c_\bk^\dag c_\bk^{}},\nonumber\\
\D_D &=& G_2 \sum_{\bk} (2\sin k_x \sin k_y) \Av{c_\bk^\dag c_\bk^{}}.
\label{eq:Nematic_Order_Definitions}
\eeqa 
Solving the above mean-field equation, Eq.~(\ref{eq:Nematic_Order_Definitions}), 
one can obtain the behavior of nematic order parameters as one varies the ratio of $F_2/G_2$ 
The transition to the diagonal nematic ordered state from isotropic liquid phase occurs above 
a critical value of interaction $G_2$, and it is second order
as a function of chemical potential. The competition between the diagonal and  the parallel nematics leads to the suppression of both phases, 
while they coexist in a finite window of chemical potential.\cite{Doh06prb}
The longer range  hopping, such as $t'$, is ignored in this study, since
qualitative features are not affected by a finite $t'$.\cite{Doh06prb,Kee03prb}
\begin{figure}[floatfix]
\includegraphics[width=3in,clip]{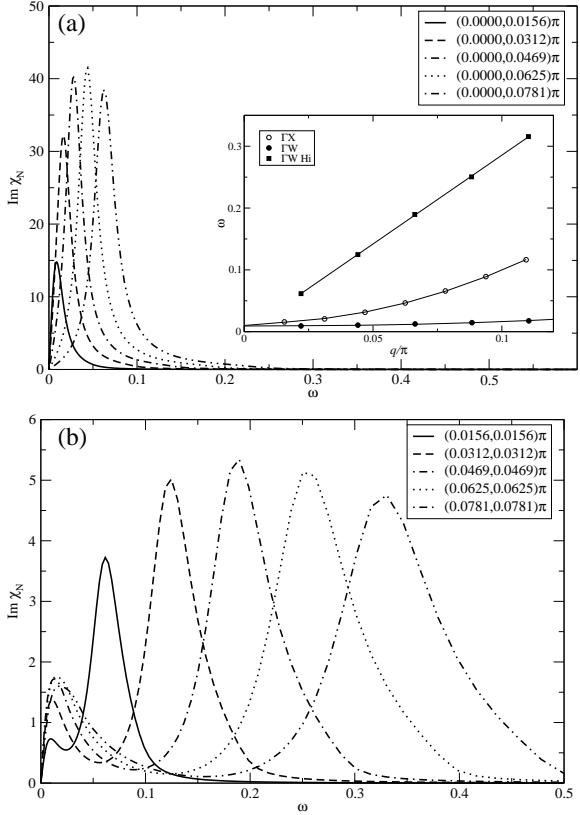}
\caption{The spectral function of the effective interaction,  $\text{Im}\,\chi_N$, as a function of $\w$ 
at slightly away from the QCP for several choices of $\bq$ along
 (a) zone parallel $(0,\pi)$ and (b) zone diagonal $(\pi,\pi)$ directions. There is a sharp propagating mode  along the
 $(0,\pi)$ direction, and the mode is diffusive along the zone diagonals. There is also a sound-mode like mode at higher energy. 
Inset: dispersion of the collective modes
along the $\Gamma W (\pi,\pi)$ (solid symbols) and $\Gamma X (0,\pi)$ (open symbols) directions. 
The lower energy modes show a $q^2$ dependence in both directions except at (sufficiently) low $q$. 
The high energy mode is sound like with dispersion $\w=v q$.
  }
\label{fig:collect}
\end{figure}

%\begin{figure}[floatfix]
%\includegraphics[width=3in,clip]{Fig2.eps}
%\caption{Dispersion of the collective mode at slightly away from the QCP
%along the $(\pi,\pi)$ (solid symbols) and $(0,\pi)$ (open symbols) directions. 
%The lower energy modes show a $q^2$ dependence in both directions except at (sufficiently) low $q$. 
%The high energy mode is sound like with dispersion $\w=v q$.
%}
%\label{fig:disp}
%\end{figure}

To study the behavior of non-Fermi liquid near the QCP, we set our parameter as follows:
$F_2N_0=0.1$ and $G_2 N_0=0.196$, where $N_0=1/(2t\pi^2)$ 
is the density of states at the Fermi surface. With these parameters, the QCP occurs
near $\mu_c/(2 t) =-0.2$. 
While the parallel nematic also occurs at smaller value of $\mu$ within the above parameter
space, below we will present the results near the $\mu_c$, 
since the non-Fermi liquid behavior is realized only 
near the QCP associated with the diagonal nematic phase.
We first consider the dynamical effective interaction between the quasi-particles.  

{\it Susceptibility in Random Phase Approximation (RPA)} ---
The effective interaction mediated by the collective modes at RPA level is given by 
\beq
\label{eq:veff}
\chi_N(\bq,\nu)= \frac{G_2(\bq)}{1-G_2(\bq)\Pi^0_N(\bq,\nu)},
\eeq
 and the bare dynamical polarizability of the diagonal nematics is given by 
\begin{eqnarray}
 \label{pid0}
 \lefteqn{\Pi^0_N(\bq,\nu) =}\nonumber\\
 &-&\int \frac{d^2p}{(2\pi)^2} \,
 \frac{f(\ve_{\bp+\bq/2}) - f(\ve_{\bp-\bq/2})}
 {\nu +i\eta- (\ve_{\bp+\bq/2} - \ve_{\bp-\bq/2})} \; 
 \z_2(\bp)^2 \; ,
\end{eqnarray}
where $f(\epsilon)$ is the Fermi function.
Near the nematic transition, the denominator in Eq.~(\ref{eq:veff}) 
becomes very small for $\bq \to 0$ and $\nu \to 0$, if $\nu$ vanishes faster than $\bq$.
This results in a singular behavior of the effective interaction in the 
forward scattering limit, and eventually non-Fermi liquid behavior.

We carried out numerical integration of the above equation. 
In  Fig.~\ref{fig:collect},
the imaginary part of the effective interaction for the diagonal nematic 
is shown for small values of $q$ for $\mu/2t=-0.22$ slightly above the QCP.
The diagonal nematic fluctuations has a propagating mode along the zone parallel directions  and 
the mode becomes diffusive along the zone diagonals ($\text{Im}\, \w/\w $ is peaked at  $\w=0$).
There exists also a sound-like mode at higher energy along the zone diagonals.  
These can be contrasted with the nematic Goldstone modes in the continuum case.\cite{Oganesyan01prb,Lawler05prb} 

In the inset of  Fig.~\ref{fig:collect}, we plot the dispersion of these modes.     
For the highly damped mode along the diagonal direction, the dispersion is determined by the peak position of the imaginary part of the effective interaction, while the mode 
dispersion along the zone parallel directions is given by the solution 
of  $1-G_2(\bq)\Pi_0(\bq,\omega(\bq))=0$. 
The lower energy modes show a $q^2$ dependence in both directions, except at very low
$q$ values with a finite but small gap of  $\Delta/2t=0.01$.  
This mode becomes soft and the effective interaction becomes singular at the QCP.
However,  the numerical integration becomes non-trivial when
$q\rightarrow 0$ where $\nu$ vanishes  faster than $q$
within the numerical computation. 

\begin{figure}[ht]
  \includegraphics[width=3in,clip]{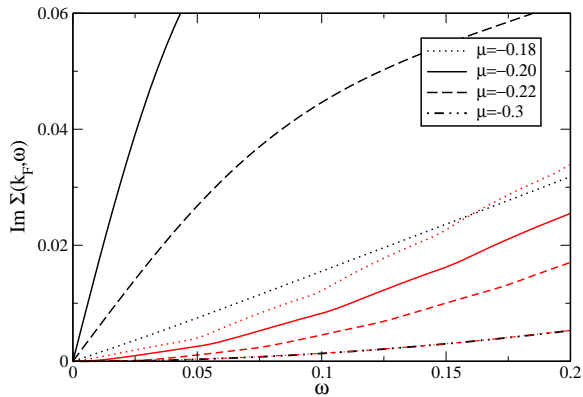}
\caption{(Color online) Calculated frequency dependence of the quasiparticle self-energy at the Fermi surface along $(\pi,\pi)$ direction at different chemical potential. 
$\mu=-0.3,-0.22$ correspond to the symmetric phase, and $\mu=-0.18$ is inside the diagonal nematic phase. $\mu=-0.2$ is the closest point to the QCP and 
shows the strongest deviation from the Fermi liquid behavior at $\w\rightarrow 0$. Non-fermi liquid behavior is observed in all cases except at $\mu=-0.3$, where  Fermi liquid behavior is restored. The red (light) curves show the contribution from the regular part of $\Pi^0_N$ to the 
self-energy. All quantities are in units of $2t$.  }
\label{fig:selfenemu}
\end{figure}
 
\textit{Quasiparticle Self-energy}-- To capture this singular behavior of the effective interaction,
we expand the $\Pi^0_N$  in the small $q$ and small $\omega/q$ limit as follows. 
\begin{equation}
\Pi^0_N\approx a + b q^2 + i c \frac{\w}{q} +O\left(\frac{\w^2}{q^2}\right)
\end{equation}
where coefficients $a,b$ and $c$  can be determined numerically,
and depend on the underlying band structure,\cite{Moriya,DellAnna06prb} and temperature. 
The closeness to the QCP can thus be quantified by quantity $\delta= 1-G_2(0) a$. 
To understand how the decay rate for single-particle excitations is modified by the collective fluctuations, we calculate the imaginary part of the self-energy.
At one-loop level, the imaginary part of the self-energy in the real frequency is given by %
\begin{eqnarray} \label{eq:imsgrpa}
\lefteqn{ \text{Im}\Si(\bk,\w) =  - \,  \int \frac{d\nu}{2 \pi} 
 \int \frac{d^2q}{(2\pi)^2} \, \z_2(\bk)\z_2(\bk+\frac{\bq}{2})\, \times}\nonumber\\
 & & \big[ b(\nu) + f(\nu+\w) \big] \, 
 \text{Im}\, \chi_N(\bq,\nu) \, \text{Im}\, G(\bk+\bq,\w+\nu) \; , 
\end{eqnarray} 
In the lowest order approximation, we use the non-interacting Green's function and  $\text{Im}\,G_0(\bk,\w)=-\pi\delta(\w-\epsilon_\bk)$ in Eq.~(\ref{eq:imsgrpa}).
Figure~\ref{fig:selfenemu} shows the imaginary parts of the self energies near the QCP  both in the symmetric phase($\mu/2t=-0.3, -0.22$) near the QCP ($\mu/2t=-0.2$) 
and inside the diagonal nematic phase ($\mu/2t=-0.18$) at the Fermi surface along the $(\pi,\pi)$ direction in the low $T$ limit. 
At small $\w$, the self-energy shows clear deviation from the Fermi liquid behavior with the asymptotic behavior $\text{Im}\Sigma \sim |\omega|^\alpha, \alpha \le 1$. 
The scattering rate is the largest close to the QCP at $\mu/2t=-0.2$, and self-energy shows a strong non-Fermi liquid behavior with 
$\alpha \sim 2/3 $\cite{Metzner03prl, DellAnna06prb}.  This indicates that the quasiparticle excitations are not well defined approaching the Fermi surface, 
which rules out a Fermi-liquid description. This non-Fermi liquid behavior exists at all parts of the Fermi surface except at four points along the $x$ and $y$ axes,
where the quasiparticle is well defined. In Fig.~\ref{fig:selfenemu}, we also compute the contribution from the regular 
part of $\Pi^0_N$ to the self-energy, and we find the amplitude is always smaller than that of the singular part. This is due to the smaller density of states at the  
Fermi surface along the nodal directions. This should be contrasted to the parallel nematic order, where the density of the states at the Fermi surface is large due to the 
closeness of the Van Hove singularity near $(\pi, 0)$.  The Fermi liquid behavior is restored far from the QCP ($\mu=-0.3$), where 
the singular and the regular parts overlap at low $\omega$. 

\begin{figure}[floatfix]
\includegraphics[width=3in,clip]{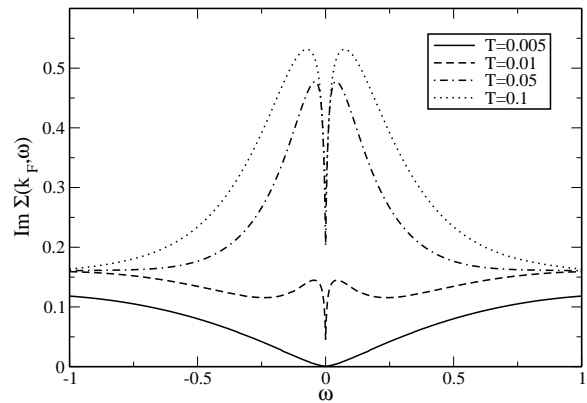}
\caption{Calculated frequency dependence of the quasiparticle self-energy at the Fermi surface along $(\pi,\pi)$ direction at different temperatures with $\mu=-0.2$. 
All quantities are in units of $2t$.}
\label{fig:selfeneT}
\end{figure}

In Fig.~\ref{fig:selfeneT}, we show the imaginary part of the self-energy at the Fermi surface in the nodal direction with $\mu/2t=-0.2$  at  different temperatures. This 
corresponds to the quantum critical regime above the QCP.\cite{Millis93prb,sachdevbook} Finite temperature effects enter through 
the finite temperature bose and fermi functions, as well as the temperature dependence of the expansion coefficients in $\Pi^0_N$. In addition to the singular contribution  of
 $\Pi^0_N$, the regular part is also included in the calculation. 
In the quantum critical regime at $T>0$, the self-energy consists both contributions from the classical and the quantum fluctuations. The quantum part of the self-energy  
obeys $(\omega/T)$ scaling in the quantum critical regime, while  the classical part dominates at $\omega=0$, and scales as 
Im $\Sigma(\mathbf{k}_F,0) \propto T\xi(T)\propto \sqrt{T/\ln T}$.\cite{Millis93prb,DellAnna06prb} We observe this scaling of the  classical fluctuations in the 
Im $\Sigma(\mathbf{k}_F,0)$; however, our numerical precision can not separate these two contributions.

%summary and conclusion
\textit{Conclusion}--- 
A consequence of the nematic order is the distortion of the underlying Fermi surface 
due to the development of the nematic phase. 
It has been proposed that the Fermi surface distortion and critical Fermi surface fluctuations play 
an important role in strongly correlated electron systems.
Recent experiments in Sr$_3$Ru$_2$O$_7$ near the metamagnetic QCP suggest that the formation of
a spin-dependent nematic phase can explain the experimental observation of the strong anisotropy 
in the magnetoresistivities.\cite{Borzi07Science,Grigera04science}
It has been also suggested  that the in-plane anisotropy of spin dynamics in detwinned 
YBa$_2$Cu$_3$O$_{7-\delta}$ have indicated a possible existence of two dimensional anisotropic liquid crystalline phase
in high temperature cuprates.\cite{Hinkov04nature,Stock04prb,Kao05prb} 
Close to the Pomeranchuk instability of the electronic system, 
the dynamics of the lattice distortion can be dramatically amplified.  
It has been argued that electron-phonon interaction indeed plays an important role 
in the high-$T_c$ superconductors.\cite{Devereaux04prl} 
The anisotropic form factors of the $B_{1g}$ buckling phonons has the form
of $(\cos{k_x}-\cos{k_y})$\cite{Devereaux04prl}. Therefore, it is interesting to note
that the effective interaction between electrons after integrating out the phonon mode
has the form of $\zeta_1({\bf k})$ in our Eq.~(\ref{eq:hamiltonian}) for the small $q$ limit.
%A further investigation on interplay between dynamics of phonon and nematic formation 
%will be an interesting subject for a future study.
While the existence of a nematic phase still need further investigation,
it is tempting to speculate that the non-Fermi liquid behaviors in these systems
can be partially attributed to the closeness
to the nematic-isotropic transition, where the Fermi surface becomes soft. 
%Angle resolved photoemission spectroscopy shows strong in-plane anisotropy in the electronic structure,
%as well as in the magnitude of the superconducting gap.\cite{Lu01prl}. 

In summary, we have studied the  fluctuation effects of the diagonal electronic nematic order in the
square lattice, using a model Hamiltonian with a quadrupole density-density interaction. 
We found that the collective mode associated with the diagonal nematic fluctuations has a strong anisotropy 
with diffusive peaks around the Fermi surface except along the $(0,\pm \pi)$ and $(\pm \pi,0)$ directions.
Close to the QCP, this mode becomes critical and strongly influences the decay rate of the quasi-particle;
the imaginary part of the self-energy shows a non-Fermi liquid behavior.  
We found the decay rate at the Fermi surface is highly anisotropic and reaches a maximum at the QCP. 
It will be interesting to explore the effects of the collective mode on charge
and heat transports, Raman spectroscopy, and electron-phonon interaction near the transition.

\begin{acknowledgments}
We thank  T. P. Devereaux, N. Nagaosa  for valuable discussions. 
This work was supported by NSC of Taiwan under grant No. NSC-94-2112-M-002-047, and Natioanl Center for Theoretical Sciences (YJK), NSERC of Canada, Canada Research
Chair, Canadian Institute for Advanced Research, and Alfred P.~Sloan Research Fellowship (HYK).  
\end{acknowledgments}

%\bibliographystyle{apsrev}
%\bibliography{./refnematic}

\end{document}